\documentclass{article}
\usepackage{spconf,amsmath,graphicx}

\usepackage[table,xcdraw]{xcolor}
\usepackage{booktabs}
\usepackage{url}

\title{Cluster-based pruning techniques for audio data}
\name{Boris Bergsma$^{1, *}$, Marta Brzezi\'{n}ska$^{1, *}$, Oleg V. Yazyev$^{1}$, Milos Cernak$^{2}$}

\address{$^{1}$Institute of Physics, Ecole Polytechnique F\'{e}d\'{e}rale de Lausanne (EPFL), \\ CH-1015 Lausanne, Switzerland \\
$^{2}$Logitech Europe S.A., Lausanne, Switzerland \\
$^{*}$equal contribution}

\begin{document}
%\ninept
%
\maketitle
\begin{abstract}

Deep learning models have become widely adopted in various domains, but their performance heavily relies on a vast amount of data. %, resulting in high computational costs.
Datasets often contain a large number of irrelevant or redundant samples, which can lead to computational inefficiencies during the training.
In this work, we introduce, for the first time in the context of the audio domain, the $k$-means clustering as a method for efficient data pruning. 
$K$-means clustering provides a way to group similar samples together, allowing the reduction of the size of the dataset while preserving its representative characteristics.
As an example, we perform clustering analysis on the keyword spotting (KWS) dataset. %to distinguish between typical and most informative samples.
%
% We then remove the defined fractions of easy and hard samples, which generates smaller datasets that we use to train three convolutional neural networks (CNNs) with $\sim$ 3k, 29k, and 270k parameters.  
% tiny ($\sim$ 3k parameters), medium ($\sim$ 29k), and large ($\sim$ 270k).
% 
We discuss how $k$-means clustering can significantly reduce the size of audio datasets while maintaining the classification performance across neural networks (NNs) with different architectures.
We further comment on the role of scaling analysis in identifying the optimal pruning strategies for a large number of samples.
Our studies serve as a proof-of-principle, demonstrating the potential of data selection with distance-based clustering algorithms for the audio domain and highlighting promising research avenues. 
%
% We highlight interesting research directions and possible extensions for the cluster-based pruning approach.
%
% In particular, we explain how our findings can be useful for small NNs which do not perform well with standard regularization techniques.

% Our findings are particularly relevant for tiny neural networks that do not perform well with standard regularization techniques, such as data augmentation, and suggest further improvements for neural networks running on low-cost microcontrollers.

\end{abstract}
%

%%%%%%%%%%%%%%%%%%%%%%
\section{Introduction}
\label{sec:introduction}
%%%%%%%%%%%%%%%%%%%%%%

With the advent of deep learning and the exponential growth in the amount of data available, the demand for efficient storage and processing has become crucial.
For instance, a recently released large language model from OpenAI -- GPT-4 -- is reportedly about six times larger than its predecessor, with one trillion hyperparameters. %~\cite{openai2023gpt4}.
Such a vast number of trainable parameters not only slows down the inference time and increases the computational cost of the model, but also requires an enormous amount of samples to train and generalize. 
Identifying the redundancies in datasets or models is therefore essential to reduce the computational demands.
In particular, discarding -- or pruning -- the irrelevant information \textit{before} the training is the most general, model-independent approach to optimize NNs.
Formally, pruning can be viewed as a discrete optimization problem, with the objective of recognizing the largest subset of irrelevant data from the training dataset while minimizing the change in network parameters resulting from removing the selected subset~\cite{yang2023dataset}.
Data pruning techniques have been used so far to improve predictions for movie ratings~\cite{saseendran2019impact} or image classification~\cite{Sorscher2022}, but not for audio data.
Within the audio domain, significant efforts have been rather devoted to model pruning using, for example, sparse architecture search and weight pruning~\cite{Sinha2020, Medhat2020, Wang2022, Martinelli22, Suedholt2023}.
%
% However, audio data pruning methods have not been so far rigorously explored. 
%
Relative to image datasets, audio datasets are generally smaller and can benefit from optimal data selection techniques for small datasets~\cite{BALKI2019_keras, bornschein2020small} or various cross-validation schemes implemented in deep learning frameworks such as Keras~\cite{chollet2015keras}.

% Additionally, optimal data selection for small datasets has been thoroughly studied~\cite{BALKI2019_keras, bornschein2020small} and various cross-validation schemes have already been implemented in popular deep learning frameworks such as Keras~\cite{chollet2015keras}.
%
%
Here, we explore a model-agnostic pruning technique based on unsupervised clustering analysis for audio data.
%
% The audio pruning approach described in this paper can be regarded as an improved version of the aforementioned techniques for handling small datasets. 
% %
% More specifically, we introduce a model-agnostic pruning technique based on unsupervised clustering analysis at the audio data level.
%
This approach allows us to reduce the size of training sets by exploiting similarities in the multidimensional feature space of a pre-trained large audio model.
Reduced datasets can then be effectively used with significantly smaller neural networks without compromising accuracy. %\MC{We could add 'a teaser' here to motivate readers to go on.} \MB{Slightly improved to get the reader's attention}
% Milos: thanks!
%
% Squeezing model sizes and training datasets is relevant for audio engineering, where current best-performing algorithms are based on hybrid DSP + NN processing, and the NN part needs to be usually very tiny to fit the target platforms. 
% %
% Full-band audio on low-power consumption devices is subject to additional constraints, with the primary focus being the user experience, which is dependent on low latency and high accuracy.
%
The goal of this paper is therefore twofold: (i) to discuss audio data pruning with $k$-means algorithm on a KWS dataset (Google Speech Commands Dataset V3 containing one-second samples with 36 classes~\cite{warden2018speech}), and (ii) to provide guidance on extending this approach to more complex datasets and different tasks.
% thus to contribute to a better understanding of the relationship between audio data pruning and other regularization methods requiring either data augmentation or model pruning.

The paper is structured as follows: Section~\ref{sec:related} introduces related work on data and model pruning, Section~\ref{sec:clustering} outlines the proposed audio data pruning method, and Section~\ref{sec:results} describes the experimental setup with obtained results. Finally, Section~\ref{sec:discussion} concludes the paper and outlines the future work.
%with a discussion and future directions.

%%%%%%%%%%%%%%%%%%%%%%
\section{Related work}
\label{sec:related}
%%%%%%%%%%%%%%%%%%%%%%

\subsection{Parameter pruning in (deep) neural networks}
Reducing the time and computational complexity of a neural network can be achieved by removing unnecessary weights, layers, or channels.
Neural network pruning can be performed at initialization, for example, with single-shot network pruning~\cite{lee2018snip}, dynamically during the training~\cite{liu2022sparse}, or iteratively to find the optimal subnetwork, such as with the lottery ticket hypothesis~\cite{frankle2019lottery}.
Another approach towards optimal data selection is to construct simpler proxy models from the target network and investigate their performance~\cite{coleman2020selection}.

\subsection{Score-based data pruning methods}
\textbf{EL2N score} Normed error-L2 distance can be used early in the training process to determine a subset of data points that carry the most information~\cite{Paul2021}.
Removing easy examples (characterized by the smallest EL2N scores) reduces the computational cost while maintaining test accuracy.
This data selection approach can be further improved by combining it with Selective-Backprop~\cite{ShenLu2023} or distillation~\cite{Sundar2023}.\\
\textbf{Forgetting score} Forgetting score tracks the number of times when a neural network misclassifies an example that it was previously classified correctly~\cite{Toneva2018, Azeemi2022}.
Samples that are rarely forgotten have a smaller impact on training accuracy and, therefore, can be pruned. 
Calculating the score is done at later stages of the training process as it requires collecting the statistics during the training.\\
\textbf{Memorization and influence estimations} Memorization is often viewed as a negative trait since it implies that a neural network cannot effectively generalize~\cite{feldman2021does, feldman2020neural}. The memorization score quantifies the increase in the probability of correctly predicting the label when an example is present in the training set compared to when it is absent.
Examples with high memorization scores are atypical and cannot be learned from the remaining data.
In addition, an influence score measures the effect of adding or removing an example from the training dataset on the test set predictions of a learned neural network.
As a result, the samples with high memorization and high influence are considered relevant.

These scoring metrics have been primarily used in image classification tasks that involve large datasets and are rather model-specific.
They necessitate iterative training to calculate the score, prune the data accordingly, and subsequently train the model from the beginning.
%
%Interestingly, Ref.~\cite{kalibhat2023measuring} studied the representation space of pretrained self-supervised models and found a subset of features with strong cross-correlations.
%
%Based on these observations, they proposed a metric called Q-score which determines if a given sample is likely to be misclassified. \MB{If time permits: check Q-score on our dataset.}
%
For example, a Q-score metric determines if a given sample will likely be misclassified based on the data's self-supervised feature representation and subsets with strong cross-correlation~\cite{kalibhat2023measuring}.
% In our work, we thus devise a model-agnostic pruning technique based on clustering analysis at the data level.

%%%%%%%%%%%%%%%%%%%%%%
\section{Methodology}
\label{sec:clustering}
%%%%%%%%%%%%%%%%%%%%%%

Standard validation techniques (such as $k$-fold cross-validation or bootstrap) often used for small datasets rely on repeated random splits of the dataset. 
%
%Our proposal is to perform more informative subsampling, as introduced in Ref.~\cite{Sorscher2022} and described in the following sections.
%
Inspired by a computer vision method~\cite{Sorscher2022}, we propose to perform more informative audio data subsampling, as described in the following sections.

\subsection{High-dimensional audio embeddings and clustering}

In image classification, the unsupervised $k$-means clustering algorithm was employed in the embedding space~\cite{Sorscher2022}
In our work, we adopt a similar clustering technique, but given the different structure of audio data, the results may not generalize in the same way.
%We adopted a similar methodology
%

%
The proposed analysis starts with a high-dimensional audio characterization -- for example, using wav2vec2 embeddings~\cite{baevski2020wav2vec} -- with the training data as input. 
The wav2vec2 model was trained to optimize a combination of two losses, namely contrastive loss and diversity loss. 
It is expected that the samples with similar characteristics (like the same word spoken by different persons or the same high tone of different musical instruments) will be close in feature space representation.
The similarities between the samples can then be quantified using the distance-based clustering algorithm.

\begin{figure}[h]
  \centering
  \centerline{\includegraphics[width=0.9\columnwidth]{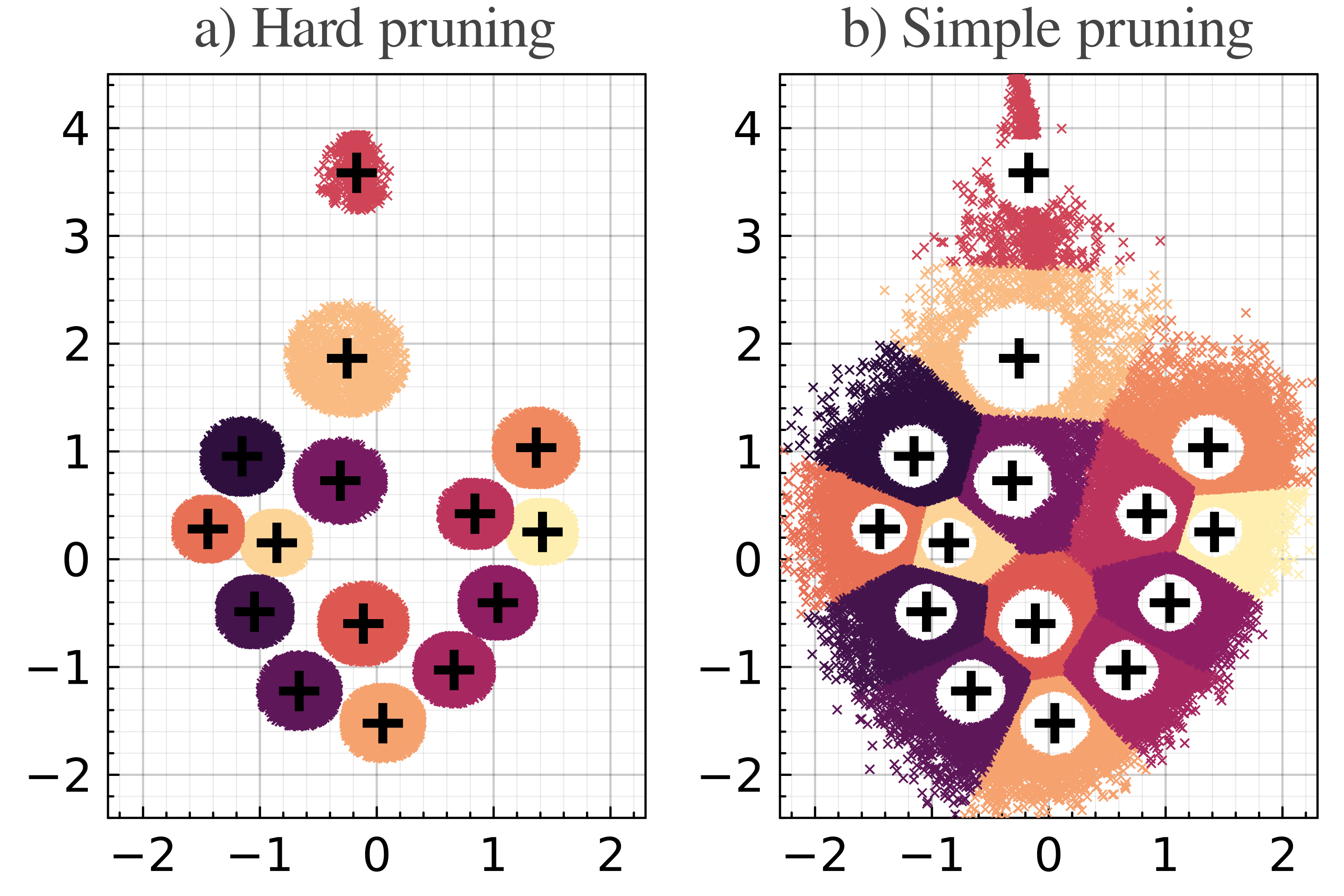}}
  \caption{Cluster-based data pruning with $k$ = 15 on a 2D projected dataset (with PCA), where we remove 50\% of (a) hard and (b) simple examples. We emphasize that our subsequent analysis solely focuses on the high-dimensional representation of the embeddings.
  }
  \label{fig:clustering}
\end{figure}

In the $k$-means clustering step, each example is represented as a point in high-dimensional space, and its proximity to a centroid is determined by the Euclidean distance metric.
Typical (hereafter denoted as \textit{simple}) samples are close to the cluster centers, whereas distinct (hereafter denoted as \textit{hard}) examples are located away from the centroids.
Therefore, we can perform simple (hard) pruning by removing the closest (farthest away) points, depending on the distance and the specified fraction of data to be kept.
To illustrate the proposed data selection approach, in Fig.~\ref{fig:clustering}, we demonstrate the $k$-means clusters obtained by projecting the data onto a 2D plane with principal component analysis (PCA).
The original formulation of $k$-means relies on the pairwise Euclidean distance.
Recently, $k$-means has been adapted to work in hyperbolic space, which is better suited for hierarchical data~\cite{chami2020trees, petermann2023hyperbolic}.
Other clustering algorithms, such as $k$-medians (based on Manhattan distance) or $k$-medoids (which minimizes arbitrary distance functions), may be more appropriate for audio data.
Therefore, a systematic analysis of clustering-based pruning with various distance metrics is needed. 
\subsection{Class selection and dataset imbalance}

The number of classes in the $k$-means algorithm is a hyperparameter that is not known a priori.
To estimate the optimal $k$, we can use PCA and compute the proportion of variance explained.
%
%The number of principal components that explain a significant amount of variance should represent the optimal value for $k$.
%
In our example, we found that a substantial number (over 150) of principal components are needed to account for around 80\% of the variance.
Importantly, performing PCA before clustering reduces the dimensionality of the data, thus accelerating the clustering analysis and data selection.
As the average complexity of the $k$-means algorithm is $\mathcal{O} (k N T)$, where $N$ stands for the number of samples and $T$ is the number of iterations, we suggest using state-of-the-art libraries (such as Faiss~\cite{johnson2019billion}) for large datasets.
%
% In Table~\ref{tab:k_scan}, we report the test accuracies for the \textit{large} model [cf. Fig.~\ref{fig:all_NNs}] depending on the number of clusters $k$.
% %
% \begin{table}[h]
% \centering
%  \resizebox{\columnwidth}{!}{%
% \begin{tabular}{c|c|c|c|c|c|c|c}
% $k$           & 25      & 45      & 65      & 85      & 105     & 155     & 250     \\ \hline
% Loss          & 0.645   & 0.634   & 0.621   & 0.620   & 0.617   & 0.615   & 0.614   \\ \hline
% Accuracy [\%] & 81.4  & 81.6  & 82.0  & 82.1  & 82.1  & 82.3  & 82.3
% \end{tabular}
% }
% \caption{Accuracies and losses (averaged over 100 runs) for different values of $k$ with a fixed pruning fraction of 20\% for simple pruning and \textit{large} model. Beyond around $k$ = 150, we do not observe any significant improvement in accuracy.}
% \label{tab:k_scan}
% \end{table}
%
%
The obtained optimal number of $k$-means classes might differ from the number of label classes, suggesting that each class might contain additional information beyond the specific keyword it represents.
We speculate that these latent features may be related to factors such as the recording equipment used, background noise, or general characteristics of the speaker (for example, gender, pitch, or speech rate).
To confirm the intuition behind these results, we propose investigating the cross-correlations within the clusters with Spearman's rank correlation. 
%

% Another important point to mention is related to dataset imbalance. 
%
Clustering algorithms often reinforce existing class imbalances, which can lead to a degradation in NN performance.
Based on the Shannon entropy, we define $balance = - \sum_i^c p_i \log (p_i) / \log (c) $, where $c$ is the number of classes and $p_i$ is the ratio of the number of elements in class $i$ to the total number of samples in the dataset.
The effect of imbalance can be countered by data augmentation.
However, augmentation creates adversarial examples, which can interfere with cluster-based data pruning.
%with or diminish the effectiveness of our
%
To understand the interplay between augmentation and pruning, it would be necessary to examine the relationship between hard and simple examples.
This can be done, for instance, by quantifying the information content of the samples (see~\cite{bergsma2022peaf, Tomar_2023_CVPR}).
%
% \begin{figure}[t]
%   \centering
%   \centerline{\includegraphics[width=\columnwidth]{WASPAA/Plots/labels_distribution_hard.pdf}}
%   \caption{Histograms of label imbalance with pruning. {\it No idea if we should even keep this plot in the final paper.}}
%   \label{fig:imbalance_pruning}
% \end{figure}
%
%%%%%%%%%%%%%%%%%%%%%%
%\section{Neural network performance for pruned datasets}
\section{Experiments and results}
%%%%%%%%%%%%%%%%%%%%%%
\label{sec:results}

The purpose of this section is to demonstrate the practical applications of cluster-based data selection described in the previous section.

\subsection{Data preparation and training}

% We use the Google Speech Commands Dataset V3 with one-second samples with 36 classes~\cite{warden2018speech}. The version provided by the \texttt{tensorflow-datasets} Python package~\cite{tf-ds} contains 95,213 samples with 5051 used for the test and another 5051 for validation sets.

%
We prepare four training sets by gradually removing simple or hard examples from the original training set (from 10\% to 40\% of data points excluded).
We then use these subsets to train three KWS systems of different sizes and measure their performance as we manipulate the number of training samples $N$.
%, denoted by $N$.
%
This allows us to investigate the asymptotic, large $N$ limit, as well as the small $N$ regime, which is relevant for devices with limited memory.
%
% \subsection{Training}
%
The audio classification is performed by extracting Mel Frequency Cepstral Coefficients (MFCCs) and passing them through a simple Convolutional Neural Network (CNN) classifier.
We constructed \textit{tiny}, \textit{small}, and \textit{large} NNs, with 3.5k, 29k and 270k parameters respectively, based on the original LeNet architecture~\cite{726791}.
%
% They are shown in Fig.~\ref{fig:all_NNs}.
% %
% \begin{figure}[t]
%   \centering
%   \centerline{\includegraphics[width=0.9\columnwidth]{WASPAA/Plots/all_nets.pdf}}
%   \caption{Neural network architectures used through the work: MFCCs followed by CNN-based classifiers. The values in the convolutional blocks (orange) represent the filter number, kernel size, and activation function.
%   The parameters in dense layers (violet) are the number of neurons and activation function.}
%   \label{fig:all_NNs}
% \end{figure}
%
% In all experiments, we used Adam optimized with an initial learning rate of $3 \cdot 10^{-4}$ for \textit{large} model and $3 \cdot 10^{-3}$ for \textit{small} and \textit{tiny} models.
% %
% We trained the networks for a maximum of 200 epochs with a batch size of 512.
% %
% However, to avoid overfitting, we employed two callbacks: \texttt{ReduceLrOnPlateau} with the patience of 3 and factor reduction set to 0.5, together with \texttt{EarlyStopping} monitoring the validation loss, with the patience of 15 and \texttt{'restore\_best\_weights' = True}.
%
%

%
We obtained the best performance of the trained models for $k$ = 155; we emphasize that the optimal value of $k$ is dataset-specific.
%
% At , we observe around five clusters per label class, suggesting that each class might contain additional information beyond the specific keyword it represents.
%
Table~\ref{tab:data_balance} presents the dataset imbalance for the simple and hard pruning methods at exemplary pruning fractions.
Notably, there is a class within our dataset that contains only the background noise, with no keywords present.
As the samples in this class have a clearly distinct structure from those with other labels, over half of them are removed by simple pruning.

\begin{table}[h]
    \centering
    \resizebox{\columnwidth}{!}{%
    \begin{tabular}{c|c|c|c|c|c}
    pruning & None & 20 \% hard & 40\% hard & 20\% simple & 40\% simple  \\
    \hline
    balance & 0.982 & 0.975 & 0.966 & 0.984 & 0.984 \\
    \end{tabular}
    }
    \caption{Dataset class imbalance for different pruning methods. A dataset with equally distributed samples is characterized by $balance = 1$. Simple pruning slightly improves the balance of the original dataset, while hard pruning leads to a more significant imbalance.}   
    \label{tab:data_balance}
 \end{table}

\subsection{Scaling analysis}

It has been observed across various domains~\cite{hestness2017deep, kaplan2020scaling} that the loss should scale as a power law with dataset size, \textit{loss} $\sim 1/N^{\nu}$, with $\nu < 1$.
While accuracy is an intuitive metric for classification tasks, it does not follow any scaling law.
To quantify the asymptotic performance of an NN, we extract the scaling exponent $\nu$ from the dependence of the loss on the number of training samples.
$\nu$ can be straightforwardly obtained from a linear regression of this dependence in a log-log scale.

% which allows us to study the performance of different architectures and data selection techniques.
%
\subsection{Results}

\begin{figure}[h]
  \centering
  \centerline{\includegraphics[width=\columnwidth]{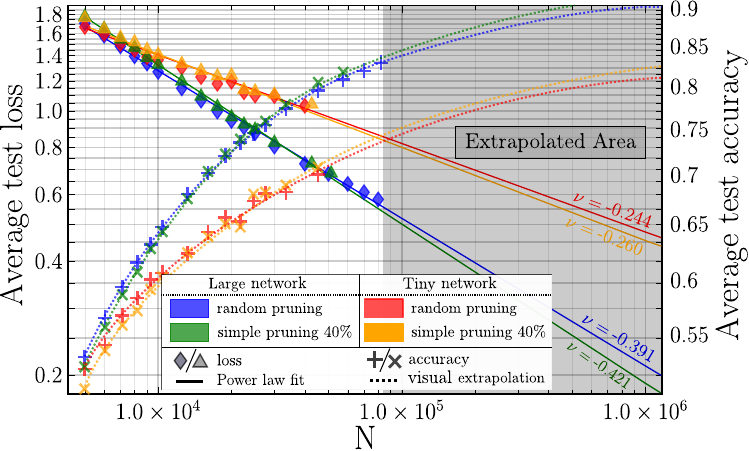}}
  \caption{Performance comparison of \textit{tiny} and \textit{large} models.
  For both architectures, we observe that simple pruning outperforms random pruning. This statement is valid for both loss and accuracy. Since the KWS dataset is rather limited, we add some extrapolation to show the potential of our method for larger datasets. The baseline Q-score pruning method performs as the random pruning and thus is not shown here.}
  \label{fig:fit_example}
\end{figure}

Fig.~\ref{fig:fit_example}, shows the test loss and test accuracy against training set size curves for the most effective pruning strategies across \textit{tiny} and \textit{large} architectures. 
The exponents $\nu$ from all experiments are collected in Table~\ref{tab:exponents}.
%
% In Fig.~\ref{fig:large_net_loss}, we show losses with respect to the $N$ for the \textit{large} neural network.
%
When the subset of training samples $N$ is smaller than the size of the entire training set, and a pruning fraction is small, it is necessary to reduce the pruned dataset further.
In such cases, we randomly remove examples from the reduced dataset.
%
% \begin{figure}[h]
%   \centering
%   \centerline{\includegraphics[width=\columnwidth]{WASPAA/Plots/Large_net_loss_plot_acc.pdf}}
%   \caption{Averaged test accuracy scaling for the \textit{large} model in a log-log scale. At small $N$, simple and hard pruning at 20 \% sample removal yield an accuracy difference of 0.3 \%. However, increasing the sample removal fraction to 40 \% results to a drop in accuracy by 1.8 \%. Above $N = 5 \cdot 10^4$, we achieve an accuracy of 80 \% for almost all pruning techniques. Hard pruning at 40 \% example removal leads to worse classification performance.}
%   \label{fig:large_net_accuracy}
% \end{figure}
%
% \begin{figure}[h]
%   \centering
%   \centerline{\includegraphics[width=\columnwidth]{WASPAA/Plots/Large_net_loss_plot.pdf}}
%   \caption{Averaged loss scaling on test set for the \textit{large} model in a log-log scale. 
%   % We quantify the effect of pruning by fitting the test loss for a given dataset and extracting the slope of the resulting line.
%   For any $N$, hard pruning at 40 \% example removal results in a larger loss. We observe that the simple pruning strategy has better asymptotic scaling.
%   % and that increasing the fraction of data removed leads to a larger scaling exponent.
%   }
%   \label{fig:large_net_loss}
% \end{figure}
%
We compare the results for various pruning strategies with random, unstructured pruning and the Q-score, informed pruning baseline~\cite{kalibhat2023measuring}.
As our datasets are relatively small, obtaining reliable performance estimates requires repeated cross-validation.
For each value of $N$, we randomly partitioned the pruned datasets into a training set 100 times and calculated the averaged scores.
This additional step is needed to demonstrate the effect of pruning clearly and does not significantly increase computational overhead.

Our analysis reveals that the scaling behavior of \textit{large} and \textit{small} models is similar, while the \textit{tiny} network performs distinctly.
Specifically, we observe that as $N$ increases, there may be either a breakdown of power-law scaling or a crossover between two regimes with different scaling exponents $\nu$. We decided to exclude these points for the exponents computations. 
In fact, the scaling with respect to the dataset size or model parameters may be a more complicated, not necessarily monotonic function.
%~\cite{caballero2023broken}.
%
\begin{table}[h]
\centering
\begin{tabular}{lccc}
\toprule
Pruning method & $\nu_{tiny}$ & $\nu_{small}$ & $\nu_{large}$ \\
\midrule
Random pruning & 0.244 & 0.396 & 0.391 \\
\midrule
10 \% hard & 0.249 & 0.390 & 0.386 \\
20 \% hard & 0.240 & 0.386 & 0.389 \\
30 \% hard & 0.239 & 0.386 & 0.388 \\
40 \% hard & 0.248 & 0.377 & 0.387 \\
\midrule
10 \% simple & 0.242 & 0.397 & 0.393 \\
20 \% simple & 0.250 & 0.406 & 0.401 \\
30 \% simple & 0.243 & 0.408 & 0.408 \\
\textbf{40 \% simple} & \textbf{0.260} & \textbf{0.413} & \textbf{0.421} \\
\bottomrule
\end{tabular}
\caption{Scaling exponents $\nu$ ($\uparrow$) obtained from $loss = f (N)$ dependencies. Larger $\nu$ indicates faster loss decay as $N$ increases. The estimated exponent for the \textit{tiny} model has an error of $0.01$, while for the \textit{small} and \textit{large} models, the error becomes smaller and is $0.005$. We find that \textit{small} and \textit{large} NNs are affected similarly by pruning; \textit{tiny} model is characterized by a smaller $\nu$. Simple pruning generally leads to more favorable scaling, while hard pruning may only benefit specific, fine-tuned cases.}
\label{tab:exponents}
\end{table}
%

%%%%%%%%%%%%%%%%%%%%%%
\section{Discussion}
\label{sec:discussion}
%%%%%%%%%%%%%%%%%%%%%%

While the differences in the exponents $\nu$ are not pronounced (see Table~\ref{tab:exponents}), our data selection method has an apparent effect on the scaling behavior:

\textbf{Simple pruning}, where we discard typical samples (close to the centroids), results in a drop in accuracy compared to random pruning when $N$ is small. As $N$ increases, the accuracy becomes comparable between these two pruning strategies. At around $N = 5 \cdot 10^4$, the accuracy for the pruned datasets surpasses that of the randomly pruned case. Additionally, we have observed that increasing the fraction of samples removed has a more noticeable impact on accuracy for \textit{small} and \textit{large} models. For the \textit{tiny} model, the 40\% pruning gives rise to the largest exponent. The obtained results of the proposed simple audio data pruning differ from the original image data pruning method~\cite{Sorscher2022}, where the authors claim that retaining easy examples (or, equivalently, removing hard) is more important for limited datasets.
However, they used massive image databases such as ImageNet to confirm their analytical findings on the knowledge distillation setting.

\textbf{Hard pruning}, where we remove most informative samples (located far from the centroids), leads to a slight overall decrease in accuracy. At the same time, the extracted scaling exponents $\nu$ are smaller. For the \textit{tiny} model, the values of $\nu$ are characterized by significant uncertainty. We verified that averaging over more training does not decrease the variance. Only the \textit{small} model exhibits a  monotonic decrease in performance as we remove more samples.
%
% \subsection{Conclusions and future work}
%

%
Generally, we found that large fractions of simple/hard pruning do not offer advantages over random pruning when $N$ is small.
This is not unexpected, as the networks may have limited exposure to diverse training data and, therefore, not generalize well.
Above $N = 5 \cdot 10^4$, we achieve an accuracy of over 80\% for almost all pruning techniques, except for hard pruning at 40\% samples removed. 
Remarkably, some classes (such as the ones corresponding to the keywords 'on' and 'no') are strongly impacted by simple pruning while remaining almost unaffected by hard pruning.
This finding suggests that a notion of similarity captured by $k$-means clustering may be close to human auditory perception.
As the $k$-means appears to identify hidden patterns, clustering may also help in improving existing data labeling.
The question of to what extent the results from the scaling analysis are transferable to other datasets remains open.

Our work is a first step towards informed cluster-based audio data pruning, and it would be of great interest to apply the proposed workflow to a larger KWS dataset such as SiDi~\cite{meneses2022sidi}, different audio processing tasks, and eventually with various audio embeddings. The implementation of the proposed method and experiments is open-sourced\footnote{\url{https://github.com/Boris-Bergsma/Audio_pruning}}.

%
% The observed discrepancy may be related to class imbalance, since hard pruning increases this imbalance (see Table~\ref{tab:data_balance}), ultimately leading to worse neural network performance.
% %

%

% %
% There are several aspects worth exploring concerning cluster-based data pruning.
%
% Firstly, it would be interesting to experiment with other clustering algorithms, such as $k$-medoids or expectation maximization.
%
% It is possible that other metrics may be better suited for the structure of audio samples.
%
% Expanding this analysis to more complex audio datasets would be a natural research direction.
% To build upon the work~\cite{bergsma2022peaf}, it would be beneficial to examine the relationship between hard and easy examples and their information content, in order to refine this data-driven pruning technique. 
%
% In particular, analyzing the correlations between hard and easy samples could lead to new insights.
%
% \section{Acknowledgement}
% \label{sec:acknowledgement}
%

\bibliographystyle{IEEEbib}
\footnotesize\bibliography{refs}

\end{document}